\documentclass{aa}  
\usepackage{natbib}
\usepackage{textcmds}
\usepackage{amsmath,amssymb}
\usepackage{graphicx}
\usepackage{color}
\usepackage{txfonts}
\begin{document}

	\title{NOEMA confirmation of an optically dark ALMA--AzTEC submillimetre galaxy at $z=5.24$}

	\subtitle{A late-stage starburst prior to quenching}
	
	\titlerunning{ASXDF1100.053.1 at $z=5.24$}
   	\author{Soh~Ikarashi
          \inst{1,2,3}, 
          R.\,J.~Ivison 
          \inst{4}, 
          William\,I.~Cowley 
          \inst{1}, 
          Kotaro~Kohno 
          \inst{5,6} 
          }
          
   	\institute{Kapteyn Astronomical Institute, University of
              Groningen, P.O.~Box 800, 9700 AV Groningen, The Netherlands
              \and
              Department of Microelectronics, Delft University of Technology,
              Postbox~5031, 2600 GA Delft, The Netherlands
              \and 
              Centre for Extragalactic Astronomy, Department of
              Physics, Durham University, South Road, Durham DH1 3LE, UK\\
              \email{soh.ikarashi@durham.ac.uk} 
              \and
              European Southern Observatory, Karl-Schwarzschild-Strasse 2, D-85748 Garching, Germany
              \and 
              Institute of Astronomy, Graduate School of Science, The University of Tokyo, 2-21-1 Osawa, Mitaka, Tokyo 181-0015, Japan
              \and 
             Research Center for the Eary Universe, School of Science, The University of Tokyo, 7-3-1 Hongo, Bunkyo, Tokyo 113-0033, Japan
             }


\abstract{We have obtained deep 1 and 3 mm spectral-line scans towards
a candidate $z\gtrsim5$ ALMA-identified AzTEC submillimetre galaxy
(SMG) in the Subaru/{\it XMM-Newton} Deep Field (or UKIDSS UDS),
ASXDF1100.053.1, using the NOrthern Extended Millimeter Array (NOEMA),
aiming to obtain its spectroscopic redshift. ASXDF1100.053.1 is an
unlensed optically dark millimetre-bright SMG with $S_{\rm 1100 \mu
m}=3.5$\,mJy and $K_{\rm AB}>25.7$ ($2\sigma$), which was expected to lie at $z=5$--7 based on its
radio--submillimetre photometric redshift. Our NOEMA spectral scan
detected line emission due to $^{12}$CO($J=5$--4) and ($J=6$--5),
providing a robust spectroscopic redshift, $z_{\rm CO}= 5.2383\pm0.0005$.
Energy-coupled spectral energy distribution (SED) modelling from
optical to radio wavelengths indicates an infrared luminosity $L_{\rm
IR}=8.3^{+1.5}_{-1.4}\times10^{12}$\,L$_{\odot}$, a star formation
rate SFR $=630^{+260}_{-380}$\,M$_{\odot}$\,yr$^{-1}$, a dust mass
$M_{\rm d}=4.4^{+0.4}_{-0.3}\times10^{8}$\,M$_{\odot}$, a stellar
mass $M_{\rm stellar}=3.5^{+3.6}_{-1.4}\times10^{11}$\,M$_{\odot}$,
and a dust temperature $T_{\rm d}=37.4^{+2.3}_{-1.8}$\,K. The CO luminosity
allows us to estimate a gas mass $M_{\rm gas}=3.1\pm
0.3\times10^{10}$\,M$_{\odot}$, suggesting a gas-to-dust mass ratio of
around 70, fairly typical for $z\sim2$ SMGs. ASXDF1100.053.1 has ALMA
continuum size $R_{\rm e}=1.0^{+0.2}_{-0.1}$\,kpc, so its surface
infrared luminosity density $\Sigma_{\rm IR}$ is
$1.2^{+0.1}_{-0.2}\times10^{12}$\,L$_{\odot}$\,kpc$^{-2}$. These physical
properties indicate that ASXDF1100.053.1 is a massive dusty
star-forming galaxy  with an unusually compact starburst. It lies close
to the star-forming main sequence at $z\sim5$, with low $M_{\rm
gas}$/$M_{\rm stellar}=0.09$, SFR/SFR$_{\rm MS} (R_{\rm SB})=0.6$, and
a gas-depletion time $\tau_{\rm dep}$ of $\approx 50$\,Myr, modulo
assumptions about the stellar initial mass function in such objects.
ASXDF1100.053.1 has extreme values of $M_{\rm gas}/M_{\rm stellar}$,
$R_{\rm SB}$, and $\tau_{\rm dep}$ compared to SMGs at $z\sim2$--4, and
those of ASXDF1100.053.1 are the smallest among SMGs at $z>5$.
ASXDF1100.053.1 is likely a late-stage dusty starburst prior to
passivisation. The number of $z=5.1$--5.3 unlensed SMGs now suggests a
number density $dN/dz=30.4\pm19.0$\,deg$^{-2}$, barely consistent
with the latest cosmological simulations.}


\maketitle
%

\section{Introduction}

When and how most massive galaxies formed in the Universe is one of
the intriguing open questions in astronomy. Since the discovery of
submillimetre galaxies \citep[SMGs;~e.g.][]{sma97,hug98,eal99} as
high-redshift dust-obscured massive star-forming galaxies with
typical infrared (IR) luminosities $L_{\rm
IR}\geq10^{12}$\,L$_{\odot}$, SMGs have been thought to be progenitors
of today's massive passive galaxies \citep[e.g.][]{san88,hop08}.
Atacama Large Millimeter/Submillimeter Array (ALMA) has shown that their far-IR dust continuum sizes are consistent
with this evolutionary link, maybe via a compact quiescent galaxy
phase \citep[e.g.][]{tof14,ika15,sim15}.

Continuum  observations at $\lambda_{\rm obs}\sim1$\,mm have a
near-uniform sensitivity to far-IR luminosity over the redshift
range $z\sim1$--6, owing to a strong negative K correction
\citep{bla93}. They are thus an ideal tool with which to study the
redshift evolution of dusty massive star-forming galaxies.
Optical--near-IR spectroscopic \citep{cha05,dan17} and photometric
redshift studies \citep[e.g.][]{mie17,cow18,sta19,dud20} of SMGs
indicate that a majority of SMGs lie at $z\sim2$--3 where the cosmic
star formation rate (SFR) is thought to peak \citep[e.g.][]{mad14,bet17,pil18,gru20}. However, there have been
suggestions that a significant fraction of SMGs could be located at
$z\gtrsim3$, especially amongst the brightest ones, which would
explain why they so often lack optical, near-IR, or radio counterparts
\citep[e.g.][]{cha05,ivi07}.

To pinpoint the positions of SMGs with an accuracy  better than 1$''$, which are originally discovered by single dishes with angular resolution of $\sim10''$--30$''$ (full width at half maximum, {\sc fwhm}), interferometric imaging is
necessary. At the highest redshifts we must move to the millimetre
(mm) or submillimetre (submm), because the  $K$ correction is not
favourable at radio wavelengths. Early submm continuum imaging of
bright SMGs using the Submillimeter Array (SMA) and the Plateau de Bure
Interferometer (PdBI) discovered a handful of radio-faint SMGs,
possibly at very high redshift \citep[e.g.][]{you07, you09, cow09,
  smo12}. We had to wait for ALMA, however, to conduct a systematic
survey of candidate $z\gtrsim4$--5 SMGs. ALMA continuum surveys towards
many hundreds of SMGs \citep[e.g.][]{hod13, ika15, ika17b, mie17,
  sta19, sim20} have pinpointed their positions and have revealed a
significant number of optically dark SMGs that are extremely faint at
optical--near-IR wavelengths \citep[e.g.][]{sim14,che15,ika15,ika17a,cow18,wil19,sma20} despite high IR
luminosities ($L_{\rm IR}=10^{12-13}$\,L$_{\odot}$) suggesting that
they are located at $z\gtrsim4$. Some $z\gtrsim4$ candidate $H$-band drop, IRAC-selected galaxies have overlapped with SMGs \citep{wan19}.  
However, the redshifts of these dark
SMGs usually remain vague. Unless they are magnified by gravitational
lensing, which then biases the resulting redshift distribution, it is
difficult and expensive to obtain spectroscopic redshifts for dark
SMGs. To do so requires an appropriately deep spectral-line scan.
Millimetre line scans towards gravitationally lensed SMGs (or dusty
star-forming galaxies, DSFGs) have been extremely successful,
revealing the redshift distribution of target lensed SMGs, including
some DSFGs at $z\gtrsim5$ \citep{vie13,wei13,reu20}. Although biased,
these results suggested strongly that the underlying submm-bright galaxy
population might have a longer high-redshift tail than had previously
been suggested. However, unambiguous confirmations (by more than two lines) of unlensed SMGs at
$z>5$ are rare: AzTEC-3 \citep{cap11,rie20}, HDF850.1 \citep{wal12}, MAMBO-9 \citep{cas19}, and GN10 \citep{rie20}. 

In the wider cosmological context, the redshifts of the optically dark SMGs are important to our understanding of the formation of massive galaxies. Recent near- and mid-IR surveys have
reported the discovery of massive, passive, and compact
quiescent galaxies (cQGs) at high redshift $z\gtrsim3$--4
\citep{str15, sch18, mer19, san19}. SMGs at even higher redshifts are
expected to be the progenitors of these massive passive galaxies, but
where are they?

To be credible, cosmological simulations must be able to generate
SMGs. Available simulations predict different redshift distributions
for SMGs, and observational updates on redshift distribution of SMGs
act to stimulate improved simulations \citep{bau05, hay13, cow15,
lac16, mca19}. Interestingly, the latest observational studies of SMGs
suggest that there may be more SMGs than previously thought
\citep{sta19,rie20,sim20}. Revealing the complete redshift
distribution of SMGs is thus essential if we are to understand the
formation of massive galaxies.

In this paper we report a pilot millimetre spectral-line scan using the
NOthern Extended Millimeter Array (NOEMA) interferometer towards a
plausible $z\gtrsim5$ candidate SMG, ASXDF1100.053.1, discovered using AzTEC and
pinpointed using ALMA \citep{ika15,ika17a}. Deep
multi-wavelength images suggest this SMG is unlikely to be lensed
\citep{ika17a,sma20}. We discuss the nature of this SMG and the
potential impact of further $z>5$ SMG surveys including our ongoing NOEMA redshift survey towards $z>5$ candidate ALMA-identified AzTEC SMGs, based on this pilot
study that explores the diversity of $z\gtrsim5$ SMGs and their volume
density. 
Despite growing evidence that the stellar initial mass
function (IMF) may be quite different in starburst environments
\citep[e.g.][]{zha18}, we adopt Chabrier's IMF \citep{cha03}
throughout this paper so that we can compare the nature of
ASXDF1100.053.1 with other systems from the literature. 
For consistency and  fair comparison, throughout this paper physical quantities of ASXDF1100.053.1 and galaxies from the literature are based on an energy-coupled spectral energy distribution (SED) model using {\sc magphys} code \citep{dac08,dac13,dac15}. 
Regarding ASXDF1100.053.1, for a consistency check between methods, some physical quantities based on classical methods are compared with those by {\sc magphys}.   
We use a cosmology with $H_{\rm 0}=70$\,km\,s$^{-1}$\,Mpc$^{-1}$,
$\Omega_{\rm M}=0.3$, and $\Omega_{\rm \Lambda}=0.7$.

\section{NOEMA scans in 1 and 3 mm spectral windows}

\subsection{Observations}

In order to determine the spectroscopic redshift of ASXDF1100.053.1,
we conducted NOEMA wide-band spectral scans.  Based on the radio--submm
photometric redshift of the target SMG, $z=6.5^{+1.4}_{-1.1}$
\citep{ika17a}, we tailored the observing plan to secure a robust
unambiguous redshift at $z\gtrsim5$.  We obtained 31\,GHz of coverage
in the 3 mm atmospheric window, targeting CO and/or [C\,{\sc i}] lines
at $z=4$--6, as well as 31\,GHz in the 1 mm window targeting the
[C\,{\sc ii}] line at $z\geq6$.

The 3 mm scan covered 80.4 to 111.4\,GHz, contiguously, via two
spectral set-ups, which made our data sensitive to at least two lines out of
CO(4--3), CO(5--4), CO(6--5), and [C\,{\sc i}](1--0) at $z\sim3$--6.  The
1 mm scan covered 240.4 to 271.4\,GHz, or $z=6.0$--6.9 in [C\,{\sc
  ii}], where the effectiveness of the 3 mm scan was expected to
deteriorate\footnote{At $z\gtrsim 6$, only high-$J$ CO transitions
  ($J\geq6$), very sensitive to CO excitation, are available in the
  3 mm window.}  at the highest redshifts. Given the known CO excitation
of SMGs \citep[e.g.][]{ivi11,rie11,bot13,spi14}, CO lines observable
at 3\,mm at $z\gtrsim6$ are expected to be very faint.  The effect of
the cosmic microwave background (CMB) on CO line flux at $z\gtrsim6$ \citep[e.g.][]{dac13,zha16} was
also a concern.  The combination of the extreme brightness of [C\,{\sc
  ii}] emission from high-redshift ultraluminous IR galaxies (ULIRGs) 
\citep[e.g.][]{sta10,swi12,coo18} and the high sky transparency at
1\,mm allows more efficient identification of redshifts at $z\gtrsim6$
than CO line scans.

Observations were conducted in 2017, 2018, and 2019 (project codes:
W17ES, W18EX, and W19EA; P.I.: Ikarashi).  The 3 mm scans were executed
using the C configuration, with ten antennas in operation, covering
baseline lengths between 24 and 370\,m.  The 1 mm scans were executed
using the C configuration with ten antennas in operation, or the D
configuration with nine antennas in operation, which covered baseline
lengths between 24 and 370\,m, and 24 and 180\,m, respectively.
3C\,454.3 or 3C\,84 were used as bandpass calibrators, LKHA\,101 or
MWC\,349 as primary flux density calibrators, and 0238$-$084 as the
local complex gain calibrator.  For one of the 3 mm tracks, 0238$-$084
was also used for bandpass calibration.  The observing log is
summarised in Table~\ref{tbl:obs}.

\begin{table*}[!h]
\begin{center}
\caption{NOEMA 3 and 1 mm spectroscopic observations.\label{tbl:obs}}
\begin{tabular}{l c c c c c c c}
\hline\hline 
Scan & Observed frequency & Observation date & On-source  &Array$^{\star}$ & Beam size$^{\dagger}$ &  Sensitivity$^{\ddagger}$  \\ 
                  & /GHz                            &                                &   time  /hrs         &       & /arcsec             &  /$\mu$Jy\,beam$^{-1}$     \\ \hline
Scan 1        &  80.4--88.15, 95.9--103.65   &     24 Feb 2019          & 3.0      &    10C     & $4.4\times1.8$, $3.8\times1.5$   & 250, 300    \\
Scan 2        &  88.15--95.9, 103.65--111.4   & 5, 7 Dec 2019    &  3.5 & 10C         &                 $4.3\times1.5$, $3.5\times1.3$    & 220, 210  \\
Scan 3        &  240.4--248.15, 255.9--263.65      &  10 Dec 2018     & 1.5                      &    10C       &                 $2.1\times0.5$, $1.9\times0.4$   & 910, 1100          \\
Scan 4        &  248.15--255.9, 263.65--271.4       &  6 Dec 2017        & 1.5                     &  9D         & $2.6\times1.3$, $2.5\times1.3$                   &   870, 970         \\
\hline
\end{tabular} \\
\end{center}
$^{\star}$ Number indicates antennas in operation during observations. 
$^{\dagger}$ Converted from visibility data with natural weighting.  
$^{\ddagger}$ Sensitivities in the 3 mm band are for a channel width of 30\,MHz, corresponding to 81--112\,km\,s$^{-1}$. Those in the 1 mm band are for a channel width of 80\,MHz, corresponding to 89--100\,km\,s$^{-1}$. 
\end{table*}

\subsection{Results}

We reduced the data using
GILDAS\footnote{http://www.iram.fr/IRAMFR/GILDAS}.  First we ran the
NOEMA pipeline for all of the data.  After the pipeline, if necessary,
we flagged additional data, corrected the phase delays, and
re-calibrated the phase and/or amplitude.  The synthesised beam size
for the 3 mm scans was $\sim4$\,arcsec ({\sc fwhm}); that for the 1 mm scans was $\sim2$--2.5\,arcsec ({\sc
  fwhm}).  The 3 mm and 1 mm datacubes achieved an r.m.s.\ noise of
210--300 and 870--1000\,$\mu$Jy\,beam$^{-1}$, respectively, for a
velocity resolution of $\sim100$\,km\,s$^{-1}$ (Table~\ref{tbl:obs}).

\begin{figure*}
    \centering
     \includegraphics[width=180mm]{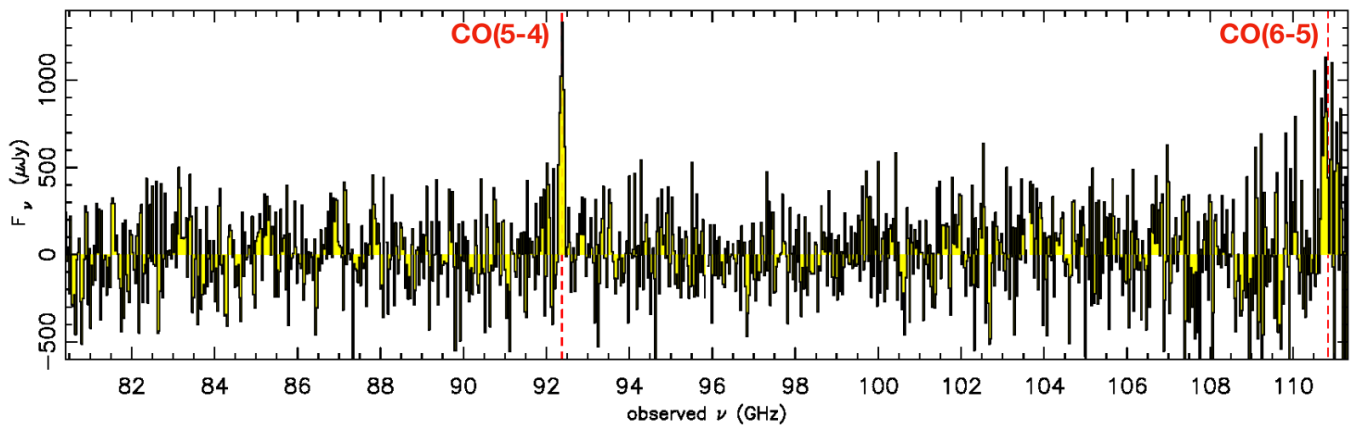}
     \caption{Composite spectrum of ASXDF1100.053.1, covering from
       80.4 to 111.3\,GHz after combining all our NOEMA
       scans at 3\,mm.  Continuum emission has been
       subtracted.  The spectral resolution is
       30\,MHz, which corresponds to 97\,km\,s$^{-1}$ and
       81\,km\,s$^{-1}$ at 92.4~GHz for CO(5--4) and at 110.8~GHz for
       CO(6--5), respectively.  }
              \label{fig:noema_spec}
\end{figure*}

\begin{figure*}
    \centering
     \includegraphics[width=180mm]{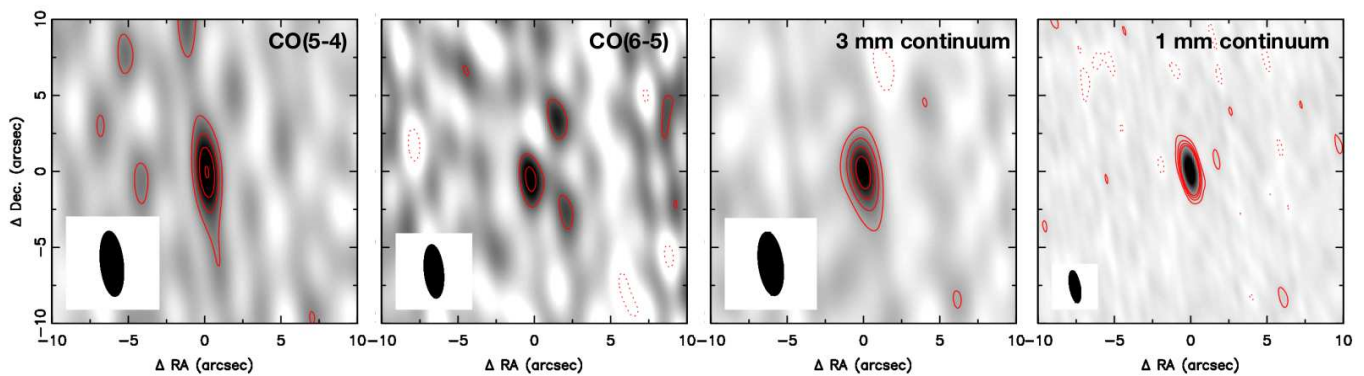}
     \caption{CO(5--4), CO(6--5), 3 mm continuum, and 1 mm continuum
       images of ASXDF1100.053.1. Contour levels are plotted at
       $\pm2.5$, 5.0, 7.5, and 10$\sigma$. The respective r.m.s.\
       levels are 79, 130, 8.9, and 63\,$\mu$Jy\,beam$^{-1}$.
       Respective synthesised beams of the CO(5--4), CO(6--5), 3 mm
       continuum, and 1 mm continuum images are 4$''$.3$\times$1$''$.5
       at a position angle (PA) of $6.0^{\circ}$,
       3$''$.5$\times$1$''$.2 (PA $5.8^{\circ}$),
       4$''$.2$\times$1$''$.6 (PA $7.7^{\circ}$), and
       2$''$.1$\times$0$''$.7 (PA $10.3^{\circ}$).  }
              \label{fig:noema_img}
\end{figure*}

We searched for line emission in the NOEMA 3 and 1 mm spectral cubes
at the position determined by ALMA for ASXDF1100.053.1
\citep{ika15,ika17a}.  In the 1 mm spectral cubes no significant
($\rm S/N\geq5$) lines were found, indicating that ASXDF1100.053.1 is not located within the [C\,{\sc ii}] redshift coverage of the 1 mm scan, $z=6.0$--6.9. In the 3 mm spectral cubes two
significant line candidates were discovered, one at
$\nu_{\rm obs}=92.37$\,GHz and the other at
$\nu_{\rm obs} = 110.83$\,GHz (Fig.~\ref{fig:noema_spec}).  Of the
potentially detectable line candidates from SMGs, the two lines
correspond to CO(5--4) and CO(6--5), suggesting that ASXDF1100.053.1
lies at $z=5.2383\pm0.0005$.  In order to evaluate the significance of the
CO(5--4) and CO(6--5) lines, we generated continuum-subtracted
CO(5--4) and CO(6--5) images by averaging channels over a width of
660\,km\,s$^{-1}$, which corresponds to $2\times2\sigma$ of $\Delta v_{\sc fwhm, \rm CO(5-4)}=380\pm58$\,km\,s$^{-1}$.  The resulting CO(5--4) and CO(6-5) maps are shown
in Fig.~\ref{fig:noema_img}.  The respective r.m.s.\ noise levels for
the CO(5--4) and CO(6--5) maps are 79 and 130\,$\mu$Jy\,beam$^{-1}$,
corresponding to 0.052 and 0.095\,Jy\,km\,s$^{-1}$\,beam$^{-1}$,
respectively.  The CO(5--4) and CO(6--5) lines are detected with S/N
values of 10 and 5.9, with $I_{\rm CO}=0.530\pm0.052$ and
$0.550\pm0.095$\, Jy\,km\,s$^{-1}$\,beam$^{-1}$, respectively.  The
emission in neither line is spatially resolved.

By averaging the spectral channels, excluding those used to make maps
of the CO lines, we have generated NOEMA  1 and 3 mm continuum maps
(Fig.~\ref{fig:noema_img}).  The continuum maps achieve r.m.s.\
sensitivities of 63 and 8.9\,$\mu$Jy\,beam$^{-1}$, respectively.
ASXDF1100.053.1 is detected with $S_{\rm 1172 \mu m}=3000\pm63$\,$\mu$Jy
(48$\sigma$) and $S_{\rm 3000 \mu m}=110\pm8.9$\,$\mu$Jy (12$\sigma$).  The
NOEMA 1 mm continuum flux density is fully consistent with our
previous measurement from ALMA,
$S_{\rm 1132 \mu m}=3450\pm100$\,$\mu$Jy \citep{ika15}, given that an
expected flux density correction factor between the NOEMA and ALMA
observing frequencies, 255 and 265\,GHz, is around 0.86 for a dust
emissivity index of $\beta=1.8$.
The observed properties are summarised  in Table~\ref{tbl:prop}. 

\section{Physical properties of ASXDF1100.053.1}

\subsection{Spectral energy distribution}\label{sec:sedana}

\begin{figure}
    \centering
     \includegraphics[width=85mm]{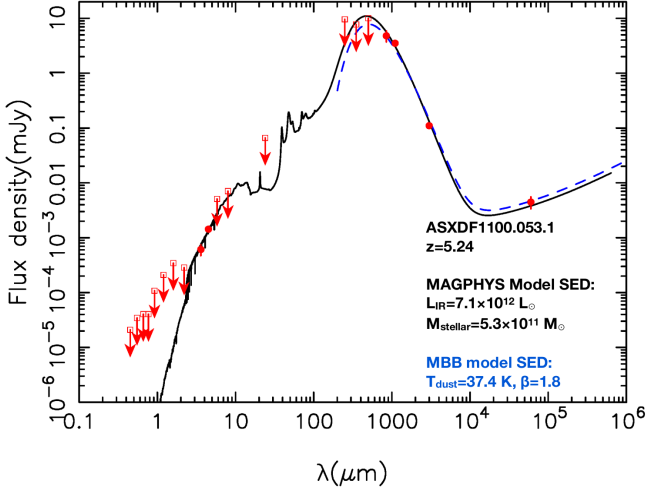}
     \caption{Spectral energy distribution  of ASXDF1100.053.1 in the observed frame. Red filled
       and open squares indicate the measured detections and upper limits,
       respectively, taken from this paper or from
       \citet{ika17a}. A solid black curve shows the best-fit SED
       obtained by our modelling using {\sc magphys} in
       \S\,\ref{sec:sedana}. 
       A dashed blue curve shows the best-fit dust SED obtained by another modelling using modified black bodies with $\beta=1.8$ including radio SED, which is presented in \S\,\ref{sec:sedana}. 
       The physical properties of the best-fit SED are not identical to the median values shown in Table~\ref{tbl:prop}.}
              \label{fig:sed}
\end{figure}

Here we derive physical properties of ASXDF1100.053.1 from
multi-wavelength data covering optical to radio wavelengths, based on
its unambiguous spectroscopic redshift.  We  modelled the spectral
energy distribution (SED) of ASXDF1100.053.1 using {\sc magphys}
\citep{dac08,dac15}.  We used Subaru SupremeCam data ($B$, $V$, $Rc$,
$i'$, and $z'$), UKIRT WFCAM data ($J$, $H$, and $K$), {\it Spitzer}
IRAC and MIPS data (3.6, 4.5, 5.8, 8.0, and 24\,$\mu$m), {\it Herschel}
SPIRE data (250, 350, and 500\,$\mu$m), SCUBA-2 850 $\mu$m data, ALMA
1100 $\mu$m data, NOEMA 3 mm data, and {\it Jansky} Very Large Array (JVLA) 6 GHz data. 
All these photometric data, except for the NOEMA 3 mm photometry, are described
in \citet{ika17a}.  We adopt the recipe for {\sc magphys} modelling
the SEDs of SMGs prescribed in \citet{dud20}, and handle the
photometric data as follows.  We treat a source as a detection if it
has at least a 3$\sigma$ detection.  For non-detections we adopt a
flux of zero and an uncertainty corresponding to 3$\sigma$ in the
UV to mid-IR bands at $\lambda_{\rm obs}\leq8$\,$\mu$m.  For
undetected sources in the IR bands at
$\lambda_{\rm obs}\geq24$\,$\mu$m, we adopt a flux density of
$1.5\pm1.0\sigma$.  Our best-fit model SED and the observed SED of
ASXDF1100.053.1 are displayed in Fig.~\ref{fig:sed}.  Physical
properties derived from the SED modelling are summarised in
Table~\ref{tbl:prop}. 
We adopt the median with the error of the 68\% range in the probability density distribution for each physical property. 

Regarding stellar mass, ASXDF1100.053.1 has significant and relevant
detections at two wavelengths, $\lambda_{\rm obs}=3.6$ and 4.5\,$\mu$m
\citep[and is tentatively detected at 5.8\,$\mu$m; see][]{ika17a}.
\citet{ika17b} estimated the stellar mass to be
$1.2^{+2.5}_{-0.81}\times10^{11}$\,M$_{\odot}$ via fits to the
UV, optical, and near-IR SED with the {\it Le Phare} code \citep{ilb06}, assuming
$z=5.5$. 
While $M_{\rm stellar}$ from {\it Le Phare} is consistent
with the $M_{\rm stellar}$ obtained from {\sc magphys} with a significance of 0.9$\sigma$, 
the offset may come from the  different assumptions for the estimation of stellar mass in these two codes \citep{mic14}; 
Among ALMA-identified SCUBA-2 850 $\mu$m SMGs in the COSMOS field, AS2COSMOS SMGs \citep{sim20}, a median $\Delta M_{\rm stellar}=M^{\rm MAGPHYS}_{\rm stellar}-M^{\rm Le Phare}_{\rm stellar}$ of 0.12$^{+0.2}_{-0.1}$\,dex for $M^{\rm Le Phare}_{\rm stellar}\geq3\times10^{10}$\,M$_{\odot}$ is found (Ikarashi et\,al., in prep) by comparing $M^{\rm MAGPHYS}_{\rm stellar}$ in the literature and $M^{\rm Le Phare}_{\rm stellar}$ by \citet{lai16}. 
In terms of absolute accuracy of stellar mass estimation of SMGs, while \citet{mic14} presented that  $M^{\rm MAGPHYS}_{\rm stellar}$ was typically higher than the intrinsic stellar mass of their model SMGs by $0.1$\,dex, 
the EAGLE simulation \citep{mca19} shows that  $M^{\rm MAGPHYS}_{\rm stellar}$ is typically lower than the intrinsic stellar mass of the model SMGs by $0.3$\,dex \citep{dud20} (see also \S\,3.4 in \citealt{hod19} for the uncertainty in stellar mass estimation of SMGs). 
The $M_{\rm stellar}$s may be
underestimated because of the lack of photometric data covering the
stellar emission at $\lambda_{\rm rest}\gtrsim1$\,$\mu$m. The
literature warns us about this potential underestimation of
$M_{\rm stellar}$, especially for massive galaxies
\citep[e.g.][]{mar06,pfo12}.

For  luminous dusty starbursts, rest-frame UV and optical wavelengths
can be contaminated by emission from active  galactic nuclei (AGN) \citep[e.g.][]{hai11,sym21},
resulting in overestimates of stellar mass. 
Extensive SED analyses of $\sim70$ ALMA-identified SMGs have revealed that half of the SMGs had non-stellar contributions of less than 10\%\    in
rest-frame $H$, and that only around 10\%\ of the SMGs had
non-stellar contribution greater than 50\%\ \citep{hai11}.  ASXDF1100.053.1
does not show any sign of a dominant AGN at UV--optical
wavelengths. We can also see from its SED at longer wavelengths that
ASXDF1100.053.1 is not a radio-loud AGN.  
Employing an AGN diagnostic using a flux ratio between  870 $\mu$m and 24 $\mu$m \citep{sta18}, the lower limit,  $\log10(F_{\rm 850\mu m}/F_{24 \mu m})>1.9$ for ASXDF1100.053.1 based on the SCUBA-2 flux and the 3$\sigma$ upper limit for MIPS 24 $\mu$m flux in \citet{ika17a}, indicates that no significant AGN contribution is expected in the IR luminosity (8--1000\,$\mu$m), which is a contribution of  less than 20\%. 
This  also indicates that a potential contribution of AGN to the IR luminosity estimated from the rest-frame far-IR SED of ASXDF1100.053.1 should be negligible, given  that in the case of a star-forming galaxy the contribution of an AGN to far-IR emission is negligible due to the contrast between the SED of an AGN and the SED of a star-forming galaxy \citep[e.g.][]{net07,kir15}. 
ASXDF1100.053.1 is thus
likely massive, in terms of stars, with
$M_{\rm stellar}\sim2$--$4\times10^{11}$\,M$_{\odot}$, though we note
that for some luminous dusty galaxies nothing is as it seems (like 
the example presented by \citealt{ivi19}), where the UV-optical SED
shows no trace of a powerful AGN lurking within the immense cocoon of
dust.

We conducted the modelling of the dust SED of ASXDF1100.053.1 via a single-component modified black-body model with dust emissivity index $\beta=1.8$ in order to estimate a classical apparent dust temperature $T^{\rm MBB}_{d}$. 
ASXDF1100.053.1  has upper limits in the photometry at $\lambda_{\rm obs}=250$, 350, and 500\,$\mu$m, where the peak of the dust SED of ASXDF1100.053.1 is expected to be located (Fig.~\ref{fig:sed}). 
For a better constraint on $T^{\rm MBB}$, we utilised the JVLA radio photometry via the radio--far-IR  luminosity correlation \citep{con92}. We modelled a radio SED by assuming a far-IR(44--122\,$\mu$m)/radio luminosity ratio $q=2.34$ in the local Universe \citep{yun01}, a SED shape of $\nu^{-0.1}$ for a thermal emission, and a SED shape of $\nu^{-0.8}$ for a non-thermal emission. 
The thermal emission and non-thermal emission are scaled so that the non-thermal emission dominates 90\% of rest-frame 1.4-GHz luminosity. 
For these assumptions, we refer to \citet{dal02}.
As a result, we obtain $T^{\rm MBB}_{d}=37.4^{+2.3}_{-1.8}$\,K and $L_{\rm FIR} (44-122 {\rm \mu m})=3.1\pm0.1\times10^{12}$\,L$_{\odot}$. This $L_{\rm FIR}$ is equivalent to $L_{\rm IR}\approx 6.2\pm0.2\times10^{12}$\,L$_{\odot}$, given the typical difference between $L_{\rm FIR}$ and $L_{\rm IR}$ \citep{bel03}. We note that the $T^{\rm MBB}$ can be cooler if there is contamination of AGN to the radio flux. 
Based on Kennicutt's conversion \citep{ken98} for \citeauthor{cha03} IMF, this $L_{\rm IR}$ from the modified black-body fitting corresponds to SFR$=620\pm20$\,M$_{\odot}$\,yr$^{-1}$, which is consistent with the SFR based on {\sc magphys}. 
In addition, we obtain $M_{\rm d}=3.6^{+1.0}_{-0.9}\times10^{8}$\,M$_{\odot}$ through the modified black-body dust modelling, adopting the equation given in \citet{hug97} and the absorption coefficient $\kappa_{\rm d}({\rm 125 \mu m})=2.64$\,m$^2$kg$^{-1}$ \citep{dun03}. This $M_{\rm d}$ from the modified black-body modelling is consistent with the $M_{\rm d}$ by {\sc magphys}.

{\sc magphys} also provides us with a dust temperature of
ASXDF1100.053.1, $T^{\sc magphys}_{d}=45.3^{+5.3}_{-4.0}$\,K. 
\citet{dud20} reported that there was a systematic offset between $T^{\sc magphys}_{\rm d}$ and the dust temperature determined by classical dust
SED modelling using a modified black body, $T^{\rm MBB}_d$.  
The $T^{\sc magphys}_{d}=45.3^{+5.3}_{-4.0}$\,K for ASXDF1100.053.1 corresponds to $T^{\rm MBB}_{\rm d}=35^{+4}_{-3}$\,K, which is consistent with the $T^{\rm MBB}_{d}=37.4^{+2.3}_{-1.8}$\,K, which we obtain by our fitting of a modified black-body SED described above.

\begin{table}[!h]
\begin{center}
\caption{Observed and physical properties of ASXDF1100.053.1  obtained in this paper. Each physical property shows the median value with the error of the 68\% range in the probability density distribution. \label{tbl:prop}}
\begin{tabular}{l l }
\hline
\multicolumn{2}{c}{Observed properties} \\
\hline
 $I_{\rm CO(5-4)}$ & $0.530\pm0.052$\,Jy\,km\,s$^{-1}$\,beam$^{-1}$ \\ 
 $I_{\rm CO(6-5)}$ & $0.550\pm0.095$\,Jy\,km\,s$^{-1}$\,beam$^{-1}$   \\
 $S_{\rm 1172 \mu m}$ & $3000\pm63$\,$\mu$Jy \\
 $S_{\rm 3000 \mu m}$ & $110\pm8.9$\,$\mu$Jy \\
\hline
\multicolumn{2}{c}{Physical properties} \\ \hline
$L_{\rm IR}$& 8.3$^{+1.5}_{-1.4}\times10^{12}$\,L$_{\odot}$  \\ 
$M_{\rm stellar}$& 3.5$^{+3.6}_{-1.4}\times10^{11}$\,M$_{\odot}$ \\
$M_{\rm d}$& 4.4$^{+0.4}_{-0.3}\times10^{8}$\,M$_{\odot}$ \\
$M_{\rm gas}$& (3.1$\pm0.3$)$\times10^{10}$\,M$_{\odot}$ \\
SFR& 630$^{+260}_{-380}$\,M$_{\odot}$\,yr$^{-1}$ \\
$A_{V}$ & 4.2$^{+0.6}_{-0.4}$\,mag\\
$T^{MBB}_{d}$ &  37.4$^{+2.3}_{-1.8}$\,K\\
$\Sigma_{\rm IR}$ & 1.2$^{+0.1}_{-0.2}\times10^{12}$\,L$_{\odot}$ \,kpc$^{-2}$\\
\hline
\end{tabular} \\
\end{center}
\end{table}

\subsection{Gas mass}

We  derived a gas mass for ASXDF1100.053.1 using the NOEMA
CO(5--4) data.  For $I_{\rm CO(5-4)}=0.54\pm0.05$\,Jy\,km\,s$^{-1}$,
we find
$L'_{\rm
  CO(5-4)}=(2.1\pm0.2)\times10^{10}$\,K\,km\,s$^{-1}$\,pc$^{-2}$.  At
the same time we obtained
$\log10(L'_{\rm CO(5-4)}/L_{\rm IR})=-2.59\pm0.05$, consistent with
the linear relation between $L'_{\rm CO(5-4)}$ and $L_{\rm IR}$ for
local and $z\sim1.5$ galaxies found by \citet{dad15},
$\log10(L'_{\rm CO(5-4)}/L_{\rm IR})=-2.52\pm0.24$. This indicates
that ASXDF1100.053.1 at $z\gtrsim5$ has a similar nature to galaxies
at lower redshift, in terms of its gas and dust.  Adopting the
$L'_{\rm CO(5-4)}/L'_{\rm CO(1-0)}$ ratio of the composite CO spectral
line energy distribution (SLED) of lensed SPT galaxies \citep[][modulo
concerns about differential lensing]{spi14},
we obtained an expected
$L'_{\rm CO(1-0)}=(3.1\pm0.3)\times10^{10}$\,K\,km\,s$^{-1}$\,pc$^{-2}$.
Finally, we obtained
$M_{\rm gas}=(3.1\pm0.3)\times10^{10}\times \alpha$\,M$_{\odot}$.  Here,
$\alpha$ is a CO luminosity-to-total molecular gas mass conversion
factor.  In this paper we adopt $\alpha=1.0$, which has been widely
accepted for local and high-redshift dusty star-forming galaxies with
$L_{\rm IR}>10^{12}$\,L$_{\odot}$ \citep[e.g.][and references
therein; see also arguments in \citealt{ivi11}]{car13}.  This is
consistent with recent measurements using ALMA at high redshift for
submillimetre-selected galaxies on or above the so-called main sequence \citep{cal18}.

There is a concern that the gas mass may have been underestimated
because the gas density and excitation requirements of CO(5--4) mean
it may not trace the entire gas distribution in a galaxy
\citep[e.g.][]{ivi11}.  An independent gas mass estimate can be made
using the dust mass, assuming a gas-to-dust mass ratio of 100, derived
from CO(1--0) observations of high-redshift ALMA-identified SMGs
\citep[][see also \citealt{sco14}]{ivi11,rie11,bot13,swi14}.  We
then obtained a gas mass of $\sim4.4\times10^{10}$\,M$_{\odot}$, which
suggests that the gas mass from CO(5--4) flux has been modestly
underestimated, by around 30\%, but we note that the
significant uncertainties on all these methods are often
unappreciated.

\subsection{Comparison with known SMGs}

Firstly, we address what causes the optical, near-IR, and mid-IR faintness of
ASXDF1100.053.1 by a comparison with the $\sim700$ ALMA-identified
SCUBA-2 SMGs of \citet{dud20}.  The mass-weighted age of
ASXDF1100.053.1 is $2.9\times10^8$\,yr, shorter than the median for
the \citeauthor{dud20} SMGs, $4.6\times10^8$\,yr.  The median $V$-band
dust attenuation for the \citeauthor{dud20} sample was
$A_{V}=2.89\pm0.04$ mag, with a 16--84${\rm th}$ percentile range of
$A_{V}=1.89$--4.24\,mag, where ASXDF1100.053.1 has
$A_{V}=4.2^{+0.6}_{-0.4}$\,mag, putting it amongst the top $\sim15$\%\ in terms of reddening.  Thus, our energy-coupled SED modelling
suggests that ASXDF1100.053.1 is faint at optical, near-IR, and mid-IR
wavelengths due to extreme reddening rather than mass-weighted age.

With $T^{\rm MBB}_{\rm d}=37.4^{+2.3}_{-1.8}$\,K, ASXDF1100.053.1 is
typical for a $\approx 1$ mm-selected SMG.  Lensed SMGs at $z=2$--6
show a mean of $T_{\rm d}=37.2\pm8.2$\,K \citep{wei13}, while the
\citeauthor{dud20} SMGs at $z\sim1$--4 show a median of
$T^{\rm MBB}_{\rm d}=30.4\pm0.3$\,K with a 16--84th${\rm th}$ percentile range of
25.7--37.3\,K.  The \citeauthor{dud20} SMGs show no evolution of
$T_{\rm d}$ with redshift across $z=1$--4.  However, the dust
temperatures of SMGs with $z_{\rm spec}\geq5$ (a mixture of far-IR
luminous galaxies selected with SCUBA-2 at 0.85\,mm, with {\it
  Herschel} at 250--500\,$\mu$m, and with the South Pole Telescope at
$>1$\,mm) are typically known to be higher than those at
$z\lesssim4$ \citep[e.g.][]{ivi16}.  The \citet{rie20} compilation of
$z\geq5$ SMGs found a median value of $T_{\rm d}=52.7\pm6.7$\,K, which
they suggested has not been biased high by sample selection; instead,
\citeauthor{rie20} attribute the higher $T_{\rm d}$ to the fact that
the majority of $z\geq5$ SMGs are likely to be starbursts, where the
community seems to have largely accepted that SMGs at lower redshift
are main-sequence galaxies, albeit typically at the more luminous
extreme, with a large fraction of mergers and interactions
\citep{eng10}.  Therefore, ASXDF1100.053.1 appears to be a relatively
rare massive main-sequence star-forming galaxy at $z\geq5$.
However, the latest observational study of $z\gtrsim4$ SMGs suggests
they are also typically on the main sequence \citep{dud20}, where the
\citeauthor{rie20} results could perhaps suffer from selection effects.

The \citeauthor{dud20} SMGs have a median stellar mass
$M_{\rm stellar}=(12.6\pm0.5)\times10^{10}$\,M$_{\odot}$, with a
16--84${\rm th}$ percentile range of
5.9--22$\times10^{10}$\,M$_{\odot}$, with no evolution across
$z=1$--4, consistent with the values for AzTEC-selected SMGs in the GOODS-S and COSMOS fields \citep{yun12,
  mie17}.  The stellar mass of ASXDF1100.053.1,
$3.5^{+3.6}_{-1.4}\times10^{11}$\,M$_{\odot}$, is in the top
16${\rm th}$ percentile of the stellar mass distribution of the
\citeauthor{dud20} SMGs.  Looking in the literature for $z\geq5$ SMGs
to compare with ASXDF1100.053.1, we find
$M_{\rm stellar}=(8.9^{+0.2}_{-0.4})\times10^{10}$\,M$_{\odot}$ for AzTEC-3
\citep{mie17b},
$M_{\rm stellar}=(3.2^{+1.0}_{-1.5})\times10^{9}$\,M$_{\odot}$ for
MAMBO-9 \citep{cas19}, and
$M_{\rm stellar}=(1.2\pm0.06)\times10^{11}$\,M$_{\odot}$ for GN10
\citep{rie20}. Thus, ASXDF1100.053.1 may be the most massive of the
known SMGs at $z\geq5$. 
We again note that the stellar mass of all SMGs in this paper including these $z>5$ SMGs, the \citeauthor{dud20} SMGs, and the \citeauthor{mie17} SMGs are estimated using {\sc magphys} and assuming \citeauthor{cha03}'s IMF. 

\section{Possible late-stage post-starburst}\label{sec:Dpost}

Here we discuss the nature of the star formation in ASXDF1100.053.1,
focusing on starburstiness, specific star formation rate,
star formation rate, and gas-to-stellar mass ratio.

We start by characterising the nature of the star formation in
ASXDF1100.053.1 in the context of the main sequence of galaxies.
Adopting a main sequence that evolves with redshift \citep{spe14}, we
obtain an expected main-sequence SFR ($\rm SFR_{\rm MS}$) of
760\,M$_{\odot}$\,yr$^{-1}$ at $z=5.2$ for the stellar mass of
ASXDF1100.053.1.  
We thus obtain a starburstiness ratio
${\rm SFR/SFR}_{\rm MS,z=5.2} (R_{\rm SB})=0.58^{+0.30}_{-0.22}$, indicating that
ASXDF1100.053.1 is on the main sequence at $z=5.2$.  Even assuming no
evolution of the main sequence with redshift (i.e.\ assuming the same
main sequence as seen at $z=2)$, ASXDF1100.053.1 shows
${\rm SFR/SFR}_{\rm MS,z=2}=1.5^{+0.70}_{-0.57}$, supporting the notion that
ASXDF1100.053.1 is a main-sequence star-forming galaxy rather than a
starburst.  Given that  SMGs are thought to be amongst the dustiest, most
intensely star-forming galaxies at any redshift, this then begs the
question of whether there are   any starbursts at $z>5$ if they are defined on
the basis of distance form the main sequence.  If the most luminous
are in fact dominated by accretion-related emission from AGN, as
suspected \citep[e.g.][]{ivi98,sym21}, then it would appear unlikely.

For a more direct quantity to characterise ASXDF1100.053.1, we obtain
a specific star formation rate (sSFR) of
$1.6^{+0.90}_{-0.65}\times10^{-9}$\,yr$^{-1}$. Such galaxies are usually
considered to be on the main sequence.

Next, we focus on the gas-to-stellar mass ratio ($\mu_{\rm gas}$), and
dust continuum, in order to probe the star formation in
ASXDF1100.053.1.  Adopting $\alpha=1.0$ for this SMG, we obtain $\mu_{\rm gas}=0.09$, based on the values of $M_{\rm gas}$ and
$M_{\rm stellar}$ described earlier. 
\citet{tac18} describes the typical $\mu_{\rm gas}$ for main-sequence galaxies as a function of
redshift, of $M_{\rm stellar}$, and of sSFR.  Given the measured $M_{\rm stellar}$, the known redshift
$z=5.24$, and the expected position on the main sequence, we obtain an
expected $\mu_{\rm gas}$ of 0.5, which is approximately five times higher than the observed $\mu_{\rm gas}$. 
We note that  the median of $\mu_{\rm gas}$ for the \citeauthor{bir20} SMGs is also 0.5, indicating that the $\mu_{\rm gas}$ of ASXDF1100.053.1 is approximately five times lower than the median of the SMGs. 

We then obtain a gas-depletion
time of 50\,Myr for ASXDF1100.053.1, which is approximately four times lower than a
typical value from the latest CO survey of ALMA-identified SMGs
\citep{bir20}, in which $\alpha=1$ is adopted consistently with our estimation (210\,Myr), though such estimates for individual galaxies are always
tentative given the range of plausible values of $\alpha$ and hence
$M_{\rm gas}$, the likelihood of feedback, and the possibility that
starbursts have a radically different stellar IMF \citep[e.g.][]{zha18}
and hence wildly different SFRs, $M_{\rm stellar}$, and so on. 
The gas depletion time of ASXDF1100.053.1 is also  15 times shorter than the expected value for the \citeauthor{tac18} main sequence for the same stellar math and redshift. 

We estimate a surface IR luminosity density $\Sigma_{\rm IR}$ for
ASXDF1100.053.1 from its ALMA continuum size, where the circularised
effective radius $R_{\rm c, e}=0.17^{+0.02}_{-0.01}$\,arcsec
\citep{ika17a}, corresponding to $1.0^{+0.2}_{-0.1}$\,kpc at $z=5.24$,
meaning that
$\Sigma_{\rm
  IR}=1.2^{+0.1}_{-0.2}\times10^{12}$\,L$_{\odot}$\,kpc$^{-2}$.

\begin{figure*}
    \centering
     \includegraphics[width=160mm]{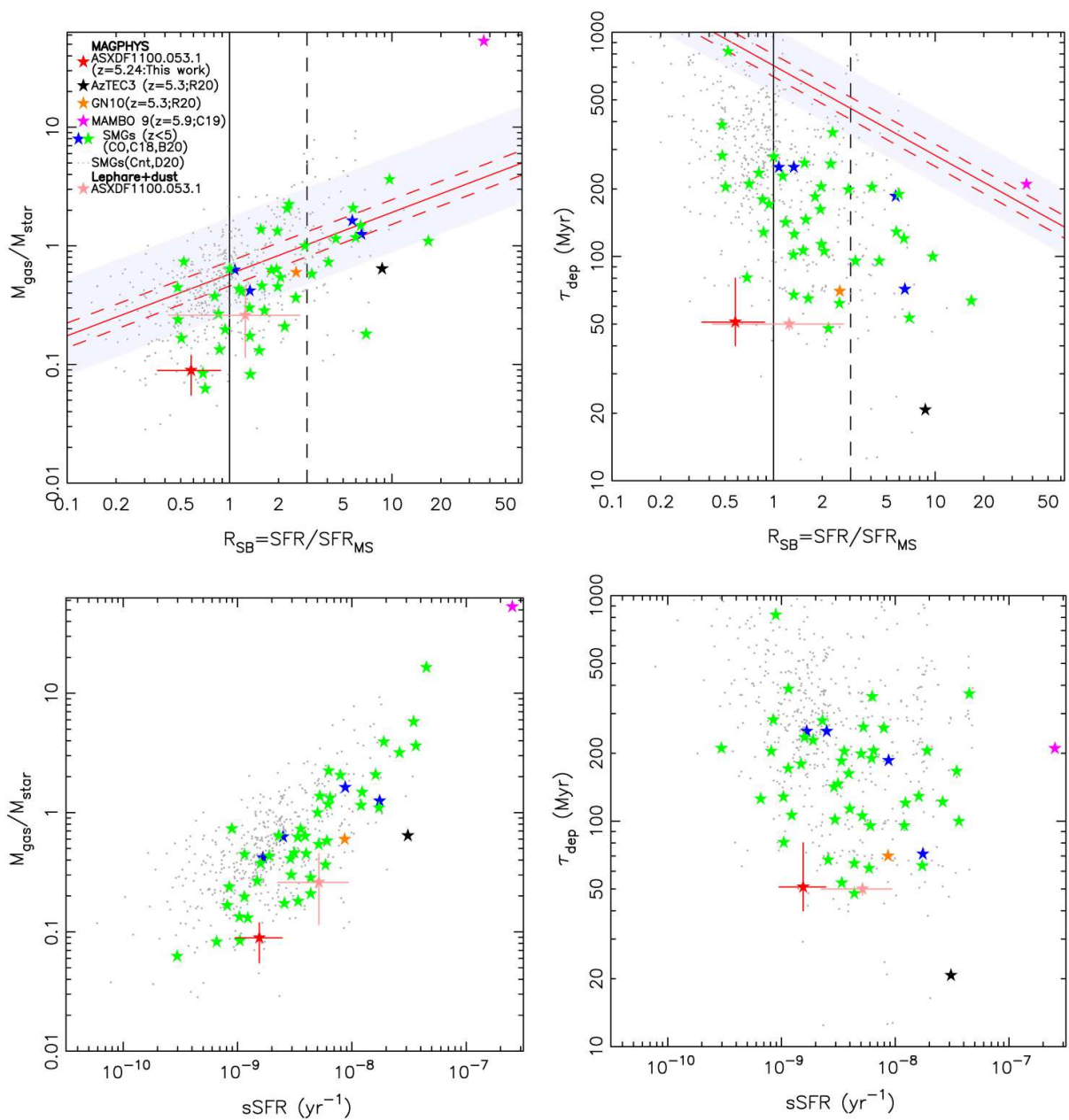}
     \caption{Indicators of the nature of star formation of ASXDF1100.053.1 to compare with other known SMGs.  {\it Top}: Gas-mass fraction
       ($\mu_{\rm gas}=M_{\rm gas}/M_{\rm stellar}$) and
       starburstiness ($R_{\rm SB}={\rm SFR/SFR}_{\rm MS}$) (left),
       and gas-depletion time ($\tau_{\rm dep}$) and starburstiness
       (right).  {\it Bottom:} Gas-mass fraction and specific
       star formation rate (sSFR) (left), and gas-depletion time and
       starburstiness (right). The red star indicates
       ASXDF1100.053.1. The black, magenta, and orange stars are the
       $z>5$ non-lensed SMGs with stellar mass estimates: AzTEC-3
       \citep{mie17b,rie20}, MAMBO-9 \citep{cas19}, and GN10
       \citep{rie20}.  The grey, blue, and green stars are  SMGs with CO
       observations at $z\sim2$--4 taken from 
       \citet{cal18} and \citet{bir20}, respectively. The small grey dots show ALMA-identified SMGs, sources at $z\sim2$--4 from
       \citet{dud20}; the gas masses of the \citeauthor{dud20} SMGs are
       derived from their dust masses assuming gas-to-dust mass ratio of
       100. The vertical solid and dashed lines mark the main sequence
       ($R_{\rm SB}=1$) and the threshold adopted for starbursts,
       $R_{\rm SB}>3$. The SFR and the $M_{\rm stellar}$ of all the SMGs in this panel to be compared with ASXDF1100.053.1 were derived using {\sc magphys}. 
       The light blue shaded area shows the relation
       for main-sequence star-forming galaxies of
       $M_{\rm stellar}=10^{10.5-11.5}$\,M$_{\odot}$ at $z=1.8$--3.4
       including $1\sigma$ scatter by the formulation in
       \citet{tac18}.  The red solid line shows the relation for
       main-sequence star-forming galaxies with
       $M_{\rm stellar}=10^{11.5}$\,M$_{\odot}$ at $z=5.2$ to be
       compared with ASXDF1100.053.1, together with expected 1$\sigma$
       scatter indicated by dashed lines. Here we adopt Chabrier's IMF \citep{cha03}
       To discuss uncertainty in stellar mass estimation, and star formation rate,  values based on stellar mass using {\it Le Phare} in \citet{ika17a}, and SFR based on the modified black-body fitting performed in \S\,\ref{sec:sedana} are indicated with a light red star with error bar. 
       }
              \label{fig:Rsb}%
\end{figure*}

To compare ASXDF1100.053.1 to other star-forming galaxies and SMGs at
lower redshift, and other SMGs at $z>5$, we show ASXDF1100.053.1 on
plots of starburstiness ($R_{\rm SB}={\rm SFR/SFR}_{\rm MS}$) versus
gas-mass fraction $\mu_{\rm gas}$ and $R_{\rm SB}$ versus
gas-depletion time $\tau_{\rm dep}$ (Fig.~\ref{fig:Rsb}).  We plot
known unlensed SMGs at $z>5$ with stellar mass estimates and gas
masses determined from CO: AzTEC-3 \citep{cap11,rie14,rie20,mie17b}, GN10 \citep{rie20}, and
MAMBO-9 \citep{cas19}; unlensed SMGs at $z\sim2$--4
\citep{ivi11,rie11,cal18,bir20}.  In addition, we add the
\citeauthor{dud20} SMGs at $z\sim2$--4, for which we derive the gas
mass from the dust mass assuming a gas-to-dust mass ratio of 100.  We
only plot $z\sim2$--4 SMGs with
$M_{\rm stellar}\geq^{10.5}$\,M$_{\odot}$.

First, we compare ASXDF1100.053.1 with other known SMGs.  While the
$\mu_{\rm gas}$ of ASXDF1100.053.1 is the lowest of the known SMGs,
the location of ASXDF1100.053.1 on the $R_{\rm SB}$--$\mu_{\rm gas}$
plot is consistent with the distribution of known SMGs; 
ASXDF1100.053.1 is located near the lowest edge of the $\mu_{\rm gas}$
and $R_{\rm SB}$ distribution of $z\sim2$--4 ALMA-identified SMGs with
CO detections \citep{cal18,bir20} and with gas masses estimated from
their dust mass \citep{dud20}.  On the plot of
$R_{\rm SB}$--$\tau_{\rm dep}$, ASXDF1100.053.1 is located at the
lowest edge of the $R_{\rm SB}$ and $\tau_{\rm dep}$ distributions of
known SMGs.  When we see sSFR instead of $R_{\rm SB}$ as a more direct
physical property, the location of ASXDF1100.053.1 on the plots of
sSFR--$\mu_{\rm gas}$ and sSFR--$\tau_{\rm dep}$ is also at the lowest
edge of these distributions of known SMGs.  This may indicate
that ASXDF1100.053.1 is closer to passivisation than other SMGs,  presumably the final phase of an SMG.

Second, we look at ASXDF1100.053.1 in the context of the other known $z>5$
unlensed SMGs.  In the literature we find three unlensed SMGs at
$z>5$ with stellar mass estimates: AzTEC-3, MAMBO-9, and GN10.  
AzTEC-3 and MAMBO-9 are located on the
opposite side from ASXDF1100.053.1 in the two plots. AzTEC-3 and
MAMBO-9 are starbursts with $R_{\rm SB}\sim10$--40.  GN10 is located in the middle of ASXDF1100.053.1,
AzTEC-3, and MAMBO-9, showing a value typical of lower-redshift SMGs.
On the $R_{\rm SB}$--$\tau_{\rm dep}$ plot, while ASXDF1100.053.1,
GN10, and AzTEC-3 all have  very low $\tau_{\rm dep}$,
ASXDF1100.053.1 shows the lowest $R_{\rm SB}$.  These plots also
suggest, therefore, that ASXDF1100.053.1 is the closest to
passivisation among the known $z>5$ SMGs.

Lastly, when we see the various physical properties of
ASXDF1100.053.1, including the low $\mu_{\rm gas}$, the short
$\tau_{\rm dep}$, and $R_{\rm SB}<3$, $R_{\rm e}=1.0$\,kpc and
$\Sigma_{\rm IR}=1.2^{+0.1}_{-0.2}\times10^{12}$\,L$_{\odot}$
\,kpc$^{-2}$, we find a similarity with the ALMA-detected
main-sequence galaxies at $z\sim2$, which are suggested to be at the
late stage of star formation prior to passivisation in \citet{elb18}.

At the bottom of this section, we briefly show how the values in Fig.~\ref{fig:Rsb} for ASXDF1100.053.1 can vary when we adopt non-energy-coupled SED modelling using stellar mass and star formation rate based on {\it Le Phare} obtained in \citet{ika17a} and the modified black-body modelling of dust emission obtained in \S\,\ref{sec:sedana}. The value for this case in each panel is displayed with a light red star in Fig.~\ref{fig:Rsb}. 
In this case, $R_{\rm SB}$ is a little bit higher than that by {\sc magphys}, and the $R_{\rm SB}$ also indicates that ASXDF1100.053.1 is on the main sequence. 
Together with this fact, considering the potential underestimation of stellar mass due to the lack of $\lambda_{\rm rest}>1$ $\mu$m photometry and the $R_{\rm SB}$ by {\sc magphys}, ASXDF1100.053.1 is very likely on the main sequence at $z=5.2$, supporting the discussion described above.

\section{Possible more massive galaxy formation at $z>5$}

Here we compare the volume density of known bright SMGs at $z>5$, 
including ASXDF1100.053.1 at $z=5.24$,  with the latest model
predictions.  We consider all known unlensed SMGs at $z=5.0$--5.5,
including\ HDF850.1 at $z=5.183$ \citep{wal12} and GN10 \citep{rie20}, both
in the Hubble Deep Field North or GOODS-N, AzTEC-3 at $z=5.298$
\citep{cap11} and ASXDF1100.053.1.  The respective flux
densities $F_{\rm 1100 \mu m}$ of ASXDF1100.053.1 and AzTEC-3 are
3.5\,mJy \citep{ika17a} and 6.2\,mJy \citep[at 1\,mm;][]{rie14}.
HDF850.1 has $F_{\rm 870 \mu m}=8.2$\,mJy \citep{cow09} and
$F_{\rm 1300 \mu m}=2.2$\,mJy \citep{dow99} and its 1100 $\mu$m
flux density is expected to be 4.3\,mJy.  GN10 has
$F_{\rm 1200 \mu m}=5.25$\,mJy and $F_{\rm 995 \mu m}=9.55$
\citep{rie20}.  Therefore, all known unlensed SMGs are $>3$\,mJy at
1100\,$\mu$m.

\begin{figure}
    \centering
     \includegraphics[width=85mm]{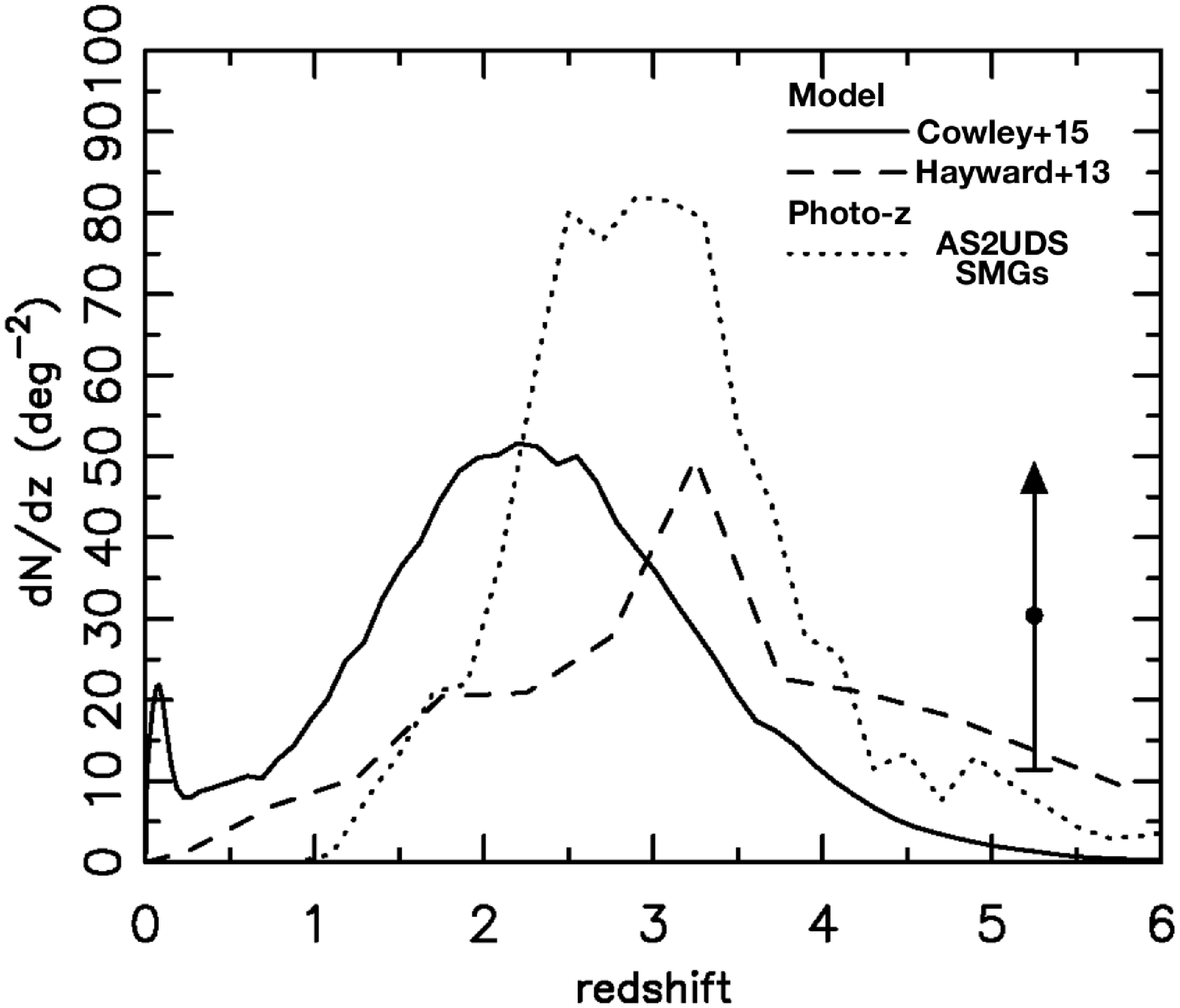}
     \caption{Redshift source number density of SMGs at $z\gtrsim5$.
       The black dot and arrow indicate a redshift source number density of
       $z=5.1$--5.3 non-lensed SMGs with 1$\sigma$ error, 
       derived from known SMGs in the literature (HDF850.1, GN10, and
       AzTEC-3) and ASXDF1100.053.1, confirmed by this study. The curved
       lines show redshift source number density distribution from
       simulations and a photometric redshift study. The solid curve
       shows a prediction for $F_{\rm 1100 \mu m}\geq3$\,mJy SMGs by
       Galform \citep{cow15}. The dashed curve shows another
       prediction for $F_{\rm 1100 \mu m}\geq3$ SMGs by
       \citet{hay13}. Here the distribution is derived by applying the
       number of $F_{\rm 1100 \mu m}\geq3$ ALMA-identified AzTEC SMGs
       \citep{ika17b} to the prediction in the simulation for flux
       density closest to the $F_{\rm 1100 \mu m}\geq3$ criterion. The
       dotted curve shows a sum of redshift probability density for
       each AS2UDS SMGs with $F_{\rm 1100 \mu m}\geq3$
       \citep{dud20}. }
              \label{fig:zdist}%
\end{figure}

Figure~\ref{fig:zdist} shows the redshift source number density along with
redshift per deg$^{-2}$ for the known (very likely) unlensed SMGs with
$F_{\rm 1100 \mu m} \geq 3$\,mJy in comparison with model predictions
for galaxies with $F_{\rm 1100 \mu m} \geq 3$\,mJy from the latest
semi-analytic cosmological model, Galform \citep{cow15,lac16} and a
semi-empirical model based on 3D hydrodynamical simulations and 3D dust
radiative transfer \citep{hay13}. 
The latest Galform model predicts the
1100 $\mu$m flux density for each mock galaxy.
\citet{hay13} predicted the redshift distribution of SMGs for
two thresholds, $F_{\rm 1100 \mu m} \geq 1$ or 4\,mJy.  We draw
the curve in Fig.~\ref{fig:zdist} by applying the model for
$F_{\rm 1100 \mu m} \geq 4$\,mJy to the observed number of
$F_{\rm 1100 \mu m} \geq 3$\,mJy ALMA-identified AzTEC SMGs
\citep{ika17b}.  In the redshift source number density calculation, we
consider a total surveyed area of 1.1 deg$^2$ for the four SMGs.  For
ASXDF1100.053.1, we adopt 1200\,arcmin$^{2}$, which is the size of the
parent AzTEC/ASTE 1100 $\mu$m map \citep{ika13}.  For HDF850.1 and
GN10, the size of the HDF-N SCUBA super-map \citep{bor03}
165\,arcmin$^{2}$ is adopted.  For AzTEC\,3, the map size of AzTEC/ASTE
1100 $\mu$m map in COSMOS, 0.72\,deg$^{2}$ \citep{are11} is
adopted.  Poissonian errors are considered for error bars.  The Galform
simulation indicates that field-to-field variations should be a minor
effect, and Poisson errors dominate in the case of SMGs
\citep{cow15}.  In addition, in order to make a comparison with the redshift
source number density of the spectroscopically confirmed $z\geq5$
SMGs, we draw the expected source number density of
$F_{\rm 1100 \mu m}\geq3$\,mJy SMGs based on the sum of photometric
redshift probability densities of the AS2UDS SMGs \citep{dud20}.  Here
we assume a constant flux ratio, $F_{\rm 870 \mu
  m}$/$F_{\rm 1100 \mu m}=1.8$.

The redshift source number density of the known unlensed bright SMGs
at $z_{\rm spec}=5.1$--5.3 is $dN/dz=30.4\pm19.0$\,deg$^{-2}$. Given  that
not all of the bright SMGs in the parent sample have a spectroscopic
redshift, this final redshift source number density can increase.  The
predicted source number densities of the latest Galform and the
Hayward models are 1.3 and 13.8, respectively; thus, the current
spectroscopically confirmed source number density is higher than these
model predictions, with significance levels of 1.5 and 0.9$\sigma$,
respectively.  As  illustrated in Fig.~\ref{fig:zdist}, recent
photometric redshift studies of hundreds of ALMA-identified SCUBA-2
SMGs reported a positive evolution of observed 870 $\mu$m flux density
with redshift \citep[][as first hinted by
\citealt{ivi07}]{sta19,sim20}, which is not seen in the Galform
model.  These photometric redshift studies and the Hayward model
suggested that bright SMGs  tend to be at higher redshift, perhaps
due to `downsizing'.  A comparison of the redshift source number
density at $z=5.1$--5.3 based on spectroscopic redshift with these
studies suggests that the redshift evolution of submillimetre flux as
a proxy of IR luminosity can evolve more strongly than these
photometric redshift studies and simulations suggested.

\section{Summary}

In order to obtain a spectroscopic redshift, we conducted wide-band
blind 1 and 3 mm scans using NOEMA to cover a contiguous 31\,GHz
block in each waveband towards the optically dark $z>5$ candidate, the
unlensed ALMA-identified AzTEC SMG with $F_{\rm 1100 \mu m}=3.5$\,mJy,
ASXDF1100.053.1.  In the NOEMA spectral scan of ASXDF1100.053.1, we
robustly detect lines of CO(5--4) and CO(6--5), showing unambiguously
that $z_{\rm spec}=5.2383\pm0.0005$.

An energy-coupled SED analysis of ASXDF1100.053.1, from optical to
radio wavelengths, then indicates that
$L_{\rm IR}=8.3^{+1.5}_{-1.4}\times10^{12}$\,L$_{\odot}$,
SFR$=630^{+260}_{-380}$\,M$_{\odot}$\,yr$^{-1}$,
$M_{\rm dust}=4.4^{+0.4}_{-0.3}\times10^{8}$\,M$_{\odot}$,
$M_{\rm stellar}=3.5^{+3.6}_{-1.4}\times10^{11}$\,M$_{\odot}$, and
$T_{\rm d}=$35$^{+4}_{-3}$\,K.  We also obtain
$M_{\rm gas}=(3.1\pm0.3)\times10^{10}$\,M$_{\odot}$ from the CO(5--4)
line luminosity.

ASXDF1100.053.1 has a low starburstiness of $R_{\rm SB}=0.58$, a low
sSFR of $1.6^{+0.90}_{-0.65}\times10^{-9}$,yr$^{-1}$, a low
gas-to-stellar mass ratio $\mu_{\rm gas}$ of 0.08, and a low
gas-depletion time ($\tau_{\rm dep}$) of 50\,Myr.  On plots of
$R_{\rm SB}$--$\mu_{\rm gas}$, $R_{\rm SB}$--$\tau_{\rm dep}$,
sSFR--$\mu_{\rm gas}$, and sSFR--$\tau_{\rm dep}$, ASXDF1100.053.1 is
located at the lowest edge of the known $z<5$ SMGs, indicating that it is
likely a late-stage dusty starburst prior to passivisation.  Among
the known unlensed $z>5$ SMGs, the location of ASXDF1100.053.1
suggests that it is the closest to passivisation.

In combination with the known unlensed $z=5.1$--5.3 SMGs, we obtain the
redshift source number density, $dN/dz=30.4\pm19.0$\,deg$^{-2}$.
Redshift confirmation of $z>5$ candidate SMGs is incomplete, so the
source number density could be higher.  Given that the latest
cosmological simulations predict $dN/dz=1$--14\,deg$^{-2}$, this
observed source number density suggests that massive galaxy formation
may have happened earlier than suggested by current models.

\begin{acknowledgements}
SI acknowledges financial support from STFC (ST/T000244/1), and the Netherlands Organization for Scientific Research (NWO) through the Top Grant Project 614.001.403 and Vidi grant No. 639.042.423. 
RIJ was funded by the Deutsche Forschungsgemeinschaft (DFG, German Research
Foundation) under Germany's Excellence Strategy -- EXC-2094 --
390783311. 
KK ackhowledges the JSPS KAKENHI Grant Number JP17H06130
and the NAOJ ALMA Scientific Research Grant Number 2017-06B. 
This work is based on observations carried out under project number W17ES, W18EX, and W19EA with the IRAM NOEMA Interferometer. IRAM is supported by INSU/CNRS (France), MPG (Germany) and IGN (Spain). 
The research leading to these results has received funding from the European Union's Horizon 2020 research and innovation program under grant agreement No 730562 [RadioNet]. 
\end{acknowledgements}

\bibliographystyle{aa} 
\bibliography{ASXDF1100.053.1_SI_v1.1.bib}

\end{document}